\documentclass{article}

\usepackage[english]{babel}
\usepackage[letterpaper,top=2cm,bottom=2cm,left=3cm,right=3cm,marginparwidth=1.75cm]{geometry}
\usepackage{amsmath}
\usepackage{graphicx}
\usepackage[colorlinks=true, allcolors=blue]{hyperref}
\usepackage{parskip}
\usepackage{booktabs}

\title{Applications of Computer Vision in Analysis of the Clock-Drawing Test as a Metric of Cognitive Impairment}
\author{Luzhou Zhang}

\date{}
\begin{document}
\maketitle

\begin{abstract}
\noindent The Clock-Drawing test is a well known and widely used neuropsychological metric to assess basic cognitive function. My objective is to combine methods of machine learning in computer vision and image analysis to predict a subject's level of cognitive impairment.
\end{abstract}

\section{Introduction}

The Clock Drawing Test (CDT) is a popular screening tool in which the subject is asked to draw a clock. In some cases, the test proctor may also request that the subject draw the clock hands to indicate a specific time, e.g. "ten minutes past 11 o'clock." This process tests several aspects of cognition, including elements of visual-spatial, planning, numerical sequencing, and motor programming abilities \cite{Kim2018}. This test is evaluated based on several metrics that represent the overall accuracy of the clock, such as the roundness of the clock face, the presence of and correct ordering of the dial numbers, and positioning of the hands to indicate the correct time \cite{Youn2021}.

Previous studies have indicated that the CDT is a good metric for detecting presence of some form of cognitive impairment (CI) \cite{Kim2018}, however, conventional testing and analysis fails to provide a reliable way to different between common causes of CI such as dementia or Alzheimer's disease. Furthermore, the manual evaluation of clock drawings is tedious and includes a level of human subjectivity that reduces its authority as a robust initial screening. \cite{Binaco2020, Ahmed2015}

To create a solution that is scalable and objective in nature, I aim to create a neural network capable of decomposing the clock drawings into their principal components and aggregating their overall accuracy scores. This unsupervised method is coupled with supervised learning where a CNN is trained on images of handdrawn clocks with attached metadata in an attempt to discriminate between types of CI.

\section{Methodology}

\subsection{Data and Images}

Data for this experiment was obtained from the National Health and Aging Trends Study (NHATS) public database, which includes around 7,500 samples of clock drawings as well as associated metadata for each image which includes salient background information on the subject as well as the position-time data for the drawing to allow the creation of a faithful simulation of the subject's speed and method for completing the drawing. 

Data for training the handwritten digit classifier was obtained from the publicly available MNIST dataset, containing 70,000 samples of handwritten digits.

\subsection{Preprocessing and Decomposition}

\begin{figure}[h]
\centering
\includegraphics[width=0.8\textwidth]{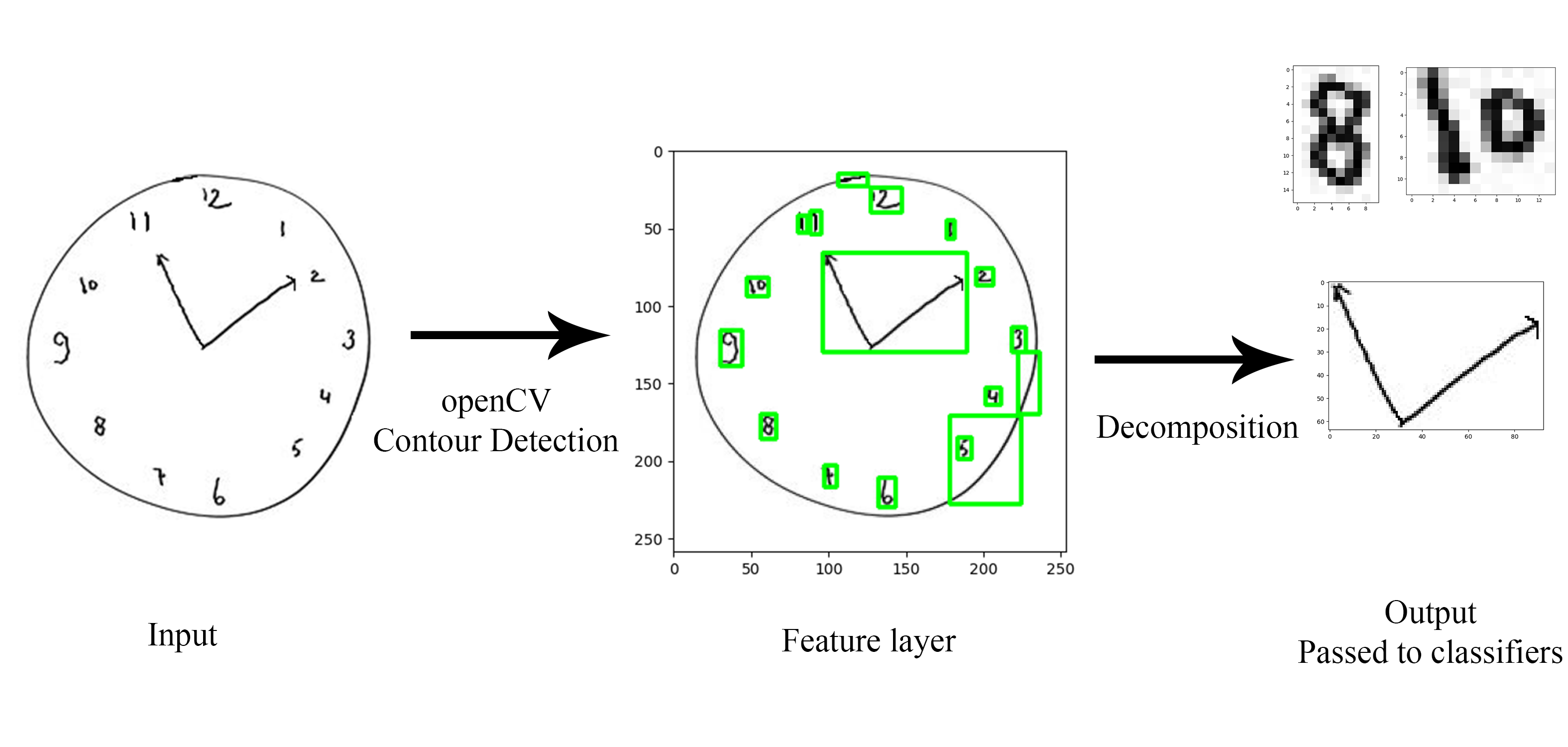}
\caption{\label{fig:cl}Process of image preprocessing and decomposition}
\end{figure}

In order to perform proper image decomposition, extensive use of contour analysis methods from the \verb|openCV-python| library were applied to the image to extract the clock face, digits, and hands. Any other non-essential details such as dial ticks were not taken into account. All methods relying on machine learning were trained and evaluated on the \verb|TensorFlow| platform for Python.

Images are first decomposed into their principal components mentioned above to reduce over-fitting and greatly increase general accuracy of all computer vision algorithms \cite{ren2022}. All salient features of the drawing are identified and separated via contour detection. The \verb|TREE| hierarchy is assigned when finding contours, enables simple extraction of the clock face by accessing the parent contour. Digits and clock hands are all first child contours. Features are grouped appropriately and passed to the downstream classifier pipeline. Non-feature contours, such as those mistakenly identified, are dropped.

7000 images were selected at random from the dataset of clock drawings. 5000 were split into the trainer set for debugging and training models. To remove any potential bias, algorithms were only run on the remaining 2000 worker set images after the model's completion.

\subsection{Digit Classification}

\begin{figure}[h]
\centering
\includegraphics[width=1\textwidth]{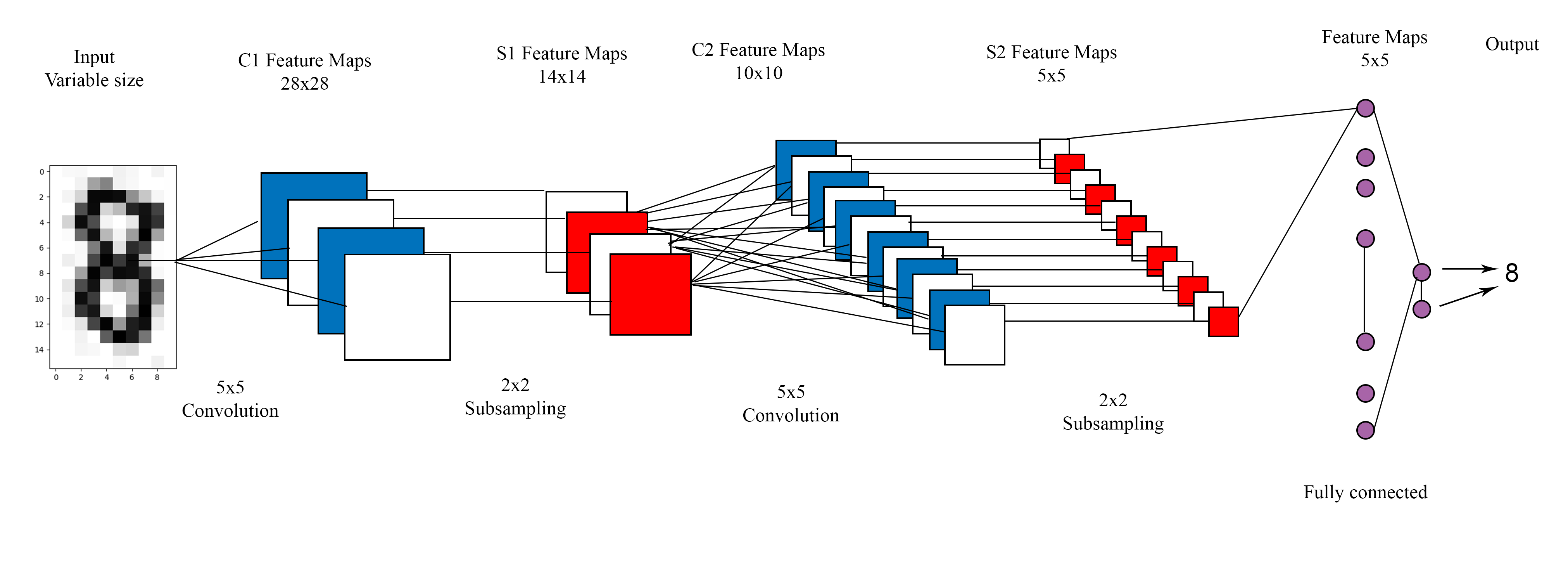}
\caption{\label{fig:frog}CNN architecture for handwritten digit extraction}
\end{figure}

To identify the handwritten digits, a CNN was trained on the MNIST database of handwritten digits, which contains a training set of 60,000 images and a test set of 10,000 images. Data preprocessing involved reshaping the images to 28x28 and applying greyscale filter and thresholding. 

Mathematically, the neuron operation is represented as
\begin{equation*} \label{eqn1}
    y[m,n]=\phi(p)=\phi(b+\sum_{k=0}^{K-1} \sum_{l=0}^{K-1} v[k,l]x[m+k,n+l])
\end{equation*}
where,
\begin{equation*} \label{eqn2}
    \phi(x)=\text{max}(0,x)
\end{equation*}
$\phi(x)$ represents the ReLU linear activator. Empirically, the sigmoidal operator is used as the activator function, but it does not provide a significant enough advantage in a larger training dataset to warrant its use over the significantly cheaper ReLU function.

The model also includes subsampling layers for data reduction and neuron output layers (containing perceptrons) to make the final classification. The equations for these are not included as they are relatively boilerplate in nature. Learning rate was optimally adjusted on a per-epoch basis via stochastic gradient descent provided by the Adam optimizer.

This CNN structure allows us the important ability to train any and all weights and biases by cycling the stochastic mode of the error back-propagation algorithm over the training data.

\subsection{Clock Face}

The clock face is rated on several metrics to grade its degree of circularity. Its centroid is determined through image moment analysis. The elliptical shape drawn can be treated as a polygon with \textit{n} vertices, where \textit{n} is sufficiently large to produce a shape that accurately models the drawn image.

Then, the center is the point $(C_{x}, C_{y})$ where
\begin{equation*}
    C_{x}=\frac{1}{6A}\sum_{i=0}^{n-1}(x_{i}+x_{i+1})(x_{i}y_{i+1}-x_{i+1}y_{i})
\end{equation*}
and
\begin{equation*}
    C_{y}=\frac{1}{6A}\sum_{i=0}^{n-1}(y_{i}+y_{i+1})(x_{i}y_{i+1}-x_{i+1}y_{i})
\end{equation*}
\textit{A} represents the signed area of the polygon as described by the shoelace formula:
\begin{equation*} \label{eqn3}
    A=\frac{1}{2}\sum_{i=0}^{n-1}(x_{i}y_{i+1}-x_{i+1}y_{i})
\end{equation*}
\begin{figure}[h]
    \centering
    \includegraphics[width=0.5\textwidth]{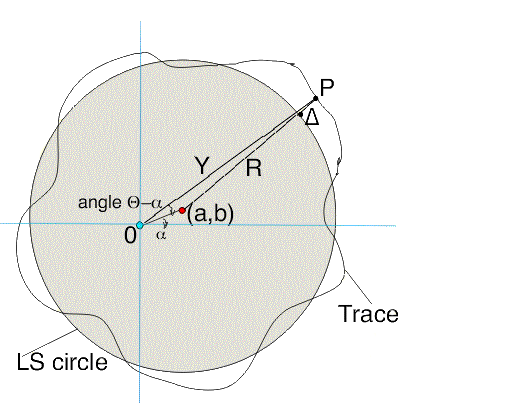}
    \caption{\label{fig:trace}Example of trace method for calculating roundness.}
\end{figure}

Roundness can be calculated by making a trace around the shape at an equal angle interval $\theta_{i}$. The radius of the trace at that specific angle is represented by $R_{i}$. A least-squares fit to the data provides the estimators detailed below:
\begin{equation*}
    \begin{gathered}
    \\
        \hat{R}=\frac{1}{N}\sum_{i=1}^{N}R_{i}\\
        \hat{a}=\frac{2}{N}\sum_{i=1}^{N}R_{i}\cos{\theta_{i}}\\
        \hat{b}=\frac{2}{N}\sum_{i=1}^{N}R_{i}\sin{\theta_{i}}
    \end{gathered}
\end{equation*}
The final graded metric is deviation, calculated as:
\begin{equation*} \label{eqn4}
    \hat{\Delta}=R_{i}-\hat{R}-\hat{a}\cos{\theta_{i}}-\hat{b}\sin{\theta_{i}}
\end{equation*}

\subsection{Clock Hands}
After edge detection is passed in preprocessing, hands can be extracted by using a randomized Hough transform to detect contours that follow a general linear shape. Using a probabilistic approach is significantly cheaper than the classic Hough transform as it avoids the expensive voting process for all candidates in the image space \cite{Stephens1991}. 

It is assumed that the hands will have contour points closest to the centroid of the clock face, so all other detected lines are ignored.

The time value that each hand represents can be calculated from its angle, where 12 o'clock is 0$^{\circ}$, 3 o'clock is 90$^{\circ}$, etc. Additionally, the angle between the hands can be used as a further evaluation metric. For the test time of 11:10, the correct angle between the hour and minute hands would be 85$^{\circ}$.

A leniency buffer of 5$^{\circ}$ is allowed to exist in any of the hand metrics, where deviation from the true value within the buffer will not be reported as a loss of accuracy.

\section{Results}
The predictor algorithm was run on 2000 worker set clock drawings and compared to results from other studies where diagnostic professionals evaluated results of CDTs. Studies were published between 1996 and 2014 and were conducted in diverse locations, including Europe, Africa, Asia, and America \cite{Park2018}. Only data for dementia CI studies were included to preserve consistency.

The classification accuracy of each study is calculated as:
\begin{equation*}
    \text{Accuracy}=\frac{\text{True Positives}+\text{True Negatives}}{\text{Total Predictions}}
\end{equation*}

\begin{table}[h]
\centering
\begin{tabular}{l c c c c c}
\toprule
& \multicolumn{5}{c}{\textbf{Comparison of CDT Evaluation Methods}} \\ \cmidrule(l){2-6}
\textbf{Study} & Location & Mean Age & Reference Standard & Scoring System & Accuracy\\
\midrule
Kirby, 2001 & Ireland & 75.2 & Neuropsychologist assessment & Sunderland & 81.7\%\\
Henderson, 2007 & UK & 59-81 & DSM IV & Manos & 79.3\%\\
Berger, 2008 & Germany & 71.5 & DSM IV & Shulman & 80.5\%\\
Chiu, 2008 & Taiwan & 75.6 & DSM IV & Rouleau & 68.9\%\\
Jouk, 2012 & Canada & 78.1 & DSM III-R & Shulman & 80.5\%\\
Ramlall, 2013 & South Africa & 75.2 & DSM IV & Rouleau & 84.3\%\\
Russo, 2014 & Argentina & 73.8 & DSM IV & Freedman & 75.6\%\\
\midrule
\midrule
Zhang, 2022 & USA & 72.0 & Empirical data & ACDTE-Zhang & 81.9\%\\
\bottomrule
\end{tabular}
\caption{CDT evaluation data}
\label{tab:resultTab}
\end{table}

\section{Conclusions}
\subsection{Discussion}
Classification accuracy of the alpha model performs within error of all other evaluation methods. Completing a run of 2000 images takes around 60 seconds. With further refinement, it is predicted that ACDTE (Automated Clock Drawing Test Evaluator) can reach $\ge$90\% accuracy, outperforming human evaluators.
\subsection{Limitations}
While precautions were taken to eliminate overfitting and proprietary training, a model that operates under this paradigm cannot be efficiently expanded to account for other evaluation parameters or variation in the drawing instructions. 

It is important to note that the CDT in itself will never be a perfect predictor of CI. Even if 100\% accuracy was achieved in a vacuum environment, real-world performance would never be fully accurate.

\subsection{Future Work}
There is room for optimization in all parts of the model and surrounding algorithms. An AI of this scale will gladly accept an increase in training complexity and time in return for increased performance and speed at runtime. As such, implementing more computationally expensive training parameters to create a more robust model, such as Canny edge detection, may prove useful.

\begin{center}
\begin{footnotesize}
\vspace{10mm}
Code and data used in this paper are available upon request.

The author declares no conflicts of interest, financial or otherwise.
\end{footnotesize}
\end{center}

\bibliographystyle{apalike}
\bibliography{main}

\end{document}